\documentclass{article}
\usepackage{spconf,amsmath,graphicx}
\usepackage{enumitem}
\setlist{nosep, leftmargin=14pt}
\let\svthefootnote\thefootnote
\newcommand\freefootnote[1]{%
  \let\thefootnote\relax%
  \footnotetext{#1}%
  \let\thefootnote\svthefootnote%
}
\usepackage[utf8]{inputenc}
\usepackage[english]{babel}
\usepackage{graphicx}

\usepackage{amsfonts} 
\usepackage{float}
\usepackage{xcolor}
\usepackage{caption}
\usepackage{subcaption}
\newcommand{\cmmnt}[1]{\ignorespaces}
\usepackage{eqnarray,amsmath}
\usepackage{amsmath,bm}
\usepackage{amssymb}
\usepackage{graphicx}
\usepackage[linesnumbered,ruled,vlined]{algorithm2e}

\title{Combining physics-based modeling and deep learning for ultrasound elastography}
\name{%
    Narges Mohammadi$^{\star}$
    \qquad Marvin M. Doyley$^{\star}$%
    \qquad Mujdat Cetin$^{\star}{}^{\dagger}$}
\address{$^{\star}$ Department of Electrical and Computer Engineering, University of Rochester, Rochester, NY, USA \\%
   $^{\dagger}$ Goergen Institute for Data Science, 
University of Rochester, Rochester, NY, USA
}
\begin{document}
\maketitle
\begin{abstract}
Ultrasound elasticity images which enable the visualization of quantitative maps of tissue stiffness can be reconstructed by solving an inverse problem. Classical model-based approaches for ultrasound elastography use deterministic finite element methods (FEMs) to incorporate the governing physical laws resulting in poor performance in noisy conditions. Moreover, these approaches utilize fixed regularizers for various tissue patterns while appropriate data-adaptive priors might be required for capturing the complex spatial elasticity distribution. 
In this regard, we propose a joint model-based and learning-based framework for estimating the elasticity distribution by solving a regularized optimization problem. We present an integrated objective function composed of a statistical physics-based forward model and a data-driven regularizer to leverage deep neural networks for learning the underlying elasticity prior. This constrained optimization problem is solved using the gradient descent (GD) method and the gradient of regularizer is simply replaced by the residual of the trained denoiser network for having an explicit objective function with reduced computation time.

\end{abstract}

\vspace{-0.2cm}
\section{Introduction}
\vspace{-0.2cm}
\freefootnote{This work has been partially supported by the National Science Foundation (NSF) under Grants CCF-1934962 and DGE-1922591.}
Elasticity imaging is concerned with the problem of reconstructing tissue parameters in terms of stiffness distribution which is the most prominent indicator of biomechanical tissue characteristics and has a large number of applications in non-invasive diagnostics and tissue property characterization.
Ultrasound elastography seeks to address cancer malignancy in various soft organs such as the liver, kidney, and lungs in a cost and time-efficient way by generating accurate and high-quality images of tissue stiffness. These non-iodized elastography methods are efficient in the early detection of soft tissue abnormalities as a replacement for tissue biopsy and can evaluate a larger area of tissue compared to a biopsy method. Elastography techniques generate the elasticity distribution by examining the tissue's response to some excitation.
Conventional FEM-based elastography methods which employ governing partial differential equations (PDEs) use Gaussian-Newton methods which assume an initial elasticity modulus and solve the constrained global stiffness equation iteratively until converging to a stationary point leading to a poor and unstable performance in noisy condition \cite{newton}. Furthermore, these methods employ fixed hand-crafted regularizers \cite{mei} for all tissue patterns although adaptive priors should be exploited for each specific tissue type. Deep neural network (DNN) capabilities propose to integrate the forward imaging model with learned data-driven priors as a constrained optimization problem rather than using end-to-end learning approaches \cite{mahsa1, tehrani, tehrani2, sima}. This infusing scheme results in both guaranteed elasticity reconstructions with the physical imaging model and exploiting complex data-driven information even with a limited training dataset \cite{willet}. \\
One group of methods for integrating physics-based modeling and deep learning priors including Plug-and-Play (PnP) \cite{pnp, narges2, carola2, narges3} and regularization by denoising (RED) \cite{RED} seeks to learn a data-adaptive denoiser and then exploits it into a regularized optimization problem as the proximal operator of regularizer or the gradient of the regularizer.  The other group of methods is based on unfolding the iterations of the minimization task in terms of a layer in a neural network including PI-GAN \cite{GAN}. Despite the PnP paradigm demonstrates successful empirical results, it does not provide any explicit representation for the objective function which yields no strong theoretical convergence guarantees to the global solution of the minimization task. On the contrary, RED methodology results in an explicit prior by replacing the regularizer gradient with denoiser network residual encouraging theoretical convergence analysis\cite{RED2}. \\
In this paper, we introduce an integrated statistical and deep learning RED  \cmmnt{iterative} paradigm for the reconstruction of elasticity modulus in noisy conditions. By this technique, the forward imaging model of ultrasound elastography is employed by a linear algebraic formulation of equilibrium equations developed with analytical error modeling of the forward model. Further, data-driven prior knowledge of the underlying elasticity distribution is learned using a DNN; and adhering to the RED scheme, the gradient of the regularizer is replaced by the residual of this denoiser network in the GD update iterations.
The resulting objective function leads to a better understanding of the solution behavior and more appropriately tuning the hyper-parameters. Our simulations implemented on a synthetic dataset manifest the improved reconstruction performance of the proposed methodology.\\
The rest of this article is arranged as follows. We describe the ultrasound elasticity imaging model and the equivalent inverse model in Sections \ref{sec2} and \ref{sec3} accordingly. The proposed RED scheme for solving the elastography inverse problem is explained in Section \ref{sec4}. The simulation results of elasticity image reconstruction are presented in Section \ref{sec5}, and at last, we provide some concluding remarks in Section \ref{sec6}. 
\vspace{-0.2cm}
\section{Forward model formulation}
\vspace{-0.1cm}
\label{sec2}
In the ultrasound imaging system, quasi-static forces on the boundary of the elastic material generate a displacement or a disturbance propagating through the medium and this dynamic response of the tissue depends upon the physical properties of the tissue. The governing physics-based model in ultrasound imaging modality is introduced by PDEs, as the system equilibrium equations, which unveil the relationship between speckle displacement and tissue elasticity \cmmnt{\cite{2DMRE}}. To have a simplified representation of the governing PDEs, finite element methods (FEM) \cite{mei0} is employed to discretize the medium cross-section.  
The resulting global stiffness equation of ultrasound elastography problem can be represented by \cite{narges2}:
\vspace{-0.1cm}
\begin{equation}
\begin{array}{l}
\label{eq:18}
\mathbf{K(E)}\mathbf{u}=\mathbf{D(u)}\mathbf{E}=\mathbf{f_{true}}\\
\end{array}
\vspace{-0.1cm}
\end{equation}
where $N$ is the number of nodes in the mesh, $\mathbf{K(E)}\in \mathbb{R}^{2N\times2N}$ indicates the global stiffness matrix, $\mathbf{D(u)}\in \mathbb{R}^{2N\times2N}$, $\mathbf{u}\in \mathbb{R}^{2N\times1}$ represents the axial and lateral displacements, $\mathbf{E}\in \mathbb{R}^{N\times1}$ is the desired elasticity distribution of the tissue over the nodes and  $\mathbf{f_{true}}\in \mathbb{R}^{2N\times1}$ indicates the Neumann boundary conditions (BCs) on observed displacement.
\vspace{-0.2cm}
\section{Inverse Problem formulation}
\label{sec3}
The statistical forward model of ultrasound elastography can be represented as follows: 
\begin{equation}
\label{eq:19-1}
\mathbf{f}=\mathbf{D(u)}\mathbf{E}+\mathbf{w}\qquad \mathbf{w}\sim \mathcal{N}(0,\,\bm{\Sigma_{w}})
\vspace{-0.1cm}
\end{equation}
where $\mathbf{f}$ stands for the noisy observed force BCs and $\mathbf{w}\in \mathbb{R}^{2N\times 1}$ indicates the Gaussian noise. 
Considering the observation process of tissue displacements $
\mathbf{u^{m}}=\mathbf{u}+\mathbf{n}$ where $\mathbf{n}\sim \mathcal{N}(0,\,\bm{\Sigma_{n}})$, $\mathbf{u^{m}}$ expresses the noisy displacement fields corrupted with noise $\mathbf{n}\in \mathbb{R}^{2N\times 1}$ with covariance matrix $\bm{\Sigma_{n}}$.
Infusing the statistical forward model in (\ref{eq:19-1}) with the noisy observation process proceeds to: 
\begin{eqnarray}
\label{eq:19-3}
\mathbf{f}&=&\mathbf{K}(\mathbf{E})\mathbf{u}+\mathbf{w}=\mathbf{K}(\mathbf{E})(\mathbf{u^{m}}-\mathbf{n})+\mathbf{w}\nonumber\\
&=&\mathbf{K}(\mathbf{E})\mathbf{u^{m}}-\mathbf{K}(\mathbf{E})\mathbf{n}+\mathbf{w}
\end{eqnarray}
Defining ${\mathbf{\Tilde{w}}}=-\mathbf{K}(\mathbf{E})\mathbf{n}+\mathbf{w}$ and utilizing noisy displacement in $\mathbf{D}(\mathbf{u^{m}})\mathbf{E}=\mathbf{K}(\mathbf{E})\mathbf{u^{m}}$ result in the following unified forward model:
\vspace{-0.2cm}
\begin{figure*}[ht]
  \centering
  \centerline{\includegraphics[width=10cm]{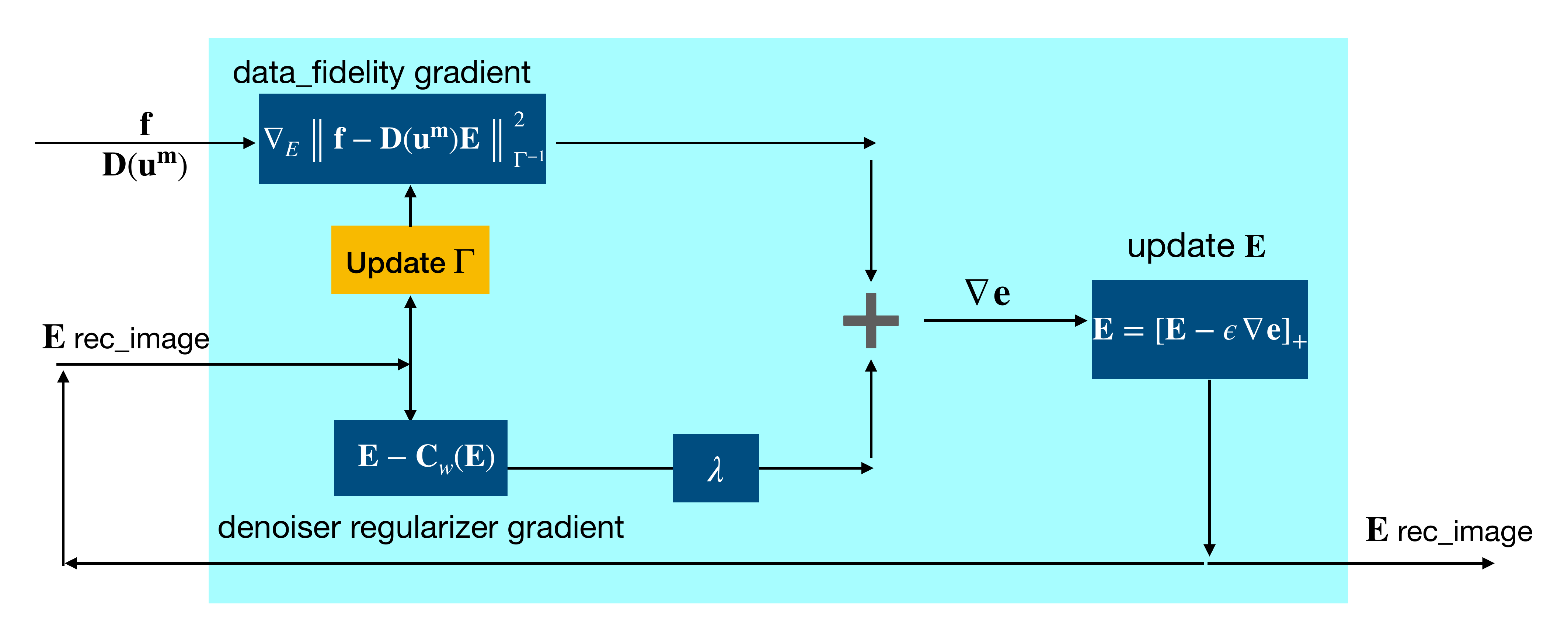}}
  \vspace{-0.45cm}
\caption{Elasticity image reconstruction using statistical physical modeling and prior learning by RED paradigm.}
\label{fig:1}
\vspace{-0.2cm}
\end{figure*}
\begin{equation}
\label{eq:19-4}
\mathbf{f}=\mathbf{D}(\mathbf{u^{m}})\mathbf{E}+\mathbf{\Tilde{w}}\qquad \mathbf{\Tilde{w}}\sim \mathcal{N}(0,\,\bm{\Gamma})
\vspace{-0.1cm}
\end{equation}
where $\bm{\Gamma}$ is denoted by:
\vspace{-0.1cm}
\begin{equation}
\label{eq:19-5}
\bm{\Gamma}=\bm{\Sigma_{w}}+\mathbf{K}(\mathbf{E})\bm{\Sigma_{n}}\mathbf{K}(\mathbf{E})^{T}
\vspace{-0.2cm}
\end{equation}
We refer to (\ref{eq:19-4}) as a statistical linear imaging model where the noise has a  signal-dependent correlated structure. By measuring $\mathbf{f}$ and $\mathbf{u^{m}}$ fields, it is required to solve a regularized optimization task for estimation of the unknown elasticity fields $\mathbf{E}$ by: 
\begin{equation}
\label{eq:20}
\begin{array}{l}
\mathbf{\hat{E}}=\mathrm{argmin} _{\mathbf{E}}\quad\frac{1}{2}\left \|  \mathbf{f}-\mathbf{D}(\mathbf{u^{m}})\mathbf{E} \right \|_{{\bm{\Gamma}}^{-1}}^{2}+\frac{N}{2}\mathrm{log}\left | \bm{\Gamma} \right |+\lambda R(\mathbf{E})\\
\quad\quad\quad
s.t.\quad \mathbf{E}>0
\end{array}
\vspace{-0.2cm}
\end{equation}
where $\left \| \mathbf{A} \right\|_{\mathbf{B}}^{2}:=(\mathbf{A}^{T}\mathbf{B}\mathbf{A})$, $R(\mathbf{E})$ expresses the regularization term and $\lambda$ is the regularization parameter. To solve this constrained optimization problem, a fixed-point paradigm \cite{fixedpoint} is employed which estimate $\mathbf{E}$ while $\bm{\Gamma}$ is fixed and this updated $\mathbf{E}$ is exploited in  (\ref{eq:19-5}) for estimating $\bm{\Gamma}$. We utilize the gradient descent (GD) as a solver to update the unknown elasticity fields $\mathbf{E}$ as follows:
\vspace{-0.2cm}
\begin{equation}
\label{eq:20-1}
\mathbf{E}\xleftarrow{}[\mathbf{E}-\epsilon (\nabla g({\mathbf{E}})+\lambda \nabla R({\mathbf{E}}))]_{+}
\vspace{-0.3cm}
\end{equation}
\vspace{-0.3cm}
where:
\begin{equation}
\label{eq:23}
g(\mathbf{E})=\frac{1}{2}(\mathbf{f}-\mathbf{D}(\mathbf{u^{m}})\mathbf{E}  )^{T}\bm{\Gamma}^{-1}(\mathbf{f}-\mathbf{D}(\mathbf{u^{m}})\mathbf{E}  )
\end{equation}
\begin{equation}
\label{eq:24}
\nabla g(\mathbf{E})=-(\mathbf{D}(\mathbf{u^{m}}))^{T}\bm{\Gamma}^{-1}(\mathbf{f}-\mathbf{D}(\mathbf{u^{m}})\mathbf{E} )
\end{equation}
$\epsilon$ denotes the solver step-size,  $[]_{+}$ expresses for the positivity condition on estimated elasticity and $\nabla R({\mathbf{E}})$ is elaborated in the following Section.
\vspace{-0.2cm}
\section{prior learning by RED paradigm}
\label{sec4}
\vspace{-0.1cm}
Following the proposed estimation procedure of elasticity $\mathbf{E}$, the prior knowledge of the underlying tissue patterns and its equivalent gradient has to be employed in (\ref{eq:20}) and (\ref{eq:20-1}). In this regard, we introduce the RED methodology for exploiting a data-adaptive regularizer for capturing the spatially variant tissue elasticity pattern. The RED scheme suggests a computationally efficient methodology for regularizer learning by a denoiser network and exploiting the residual of obtained denoiser network as the gradient of regularizer in (\ref{eq:20-1}).
RED methodology introduces the regularizer gradient by:
\begin{equation}
\label{eq:25}
\nabla R(\mathbf{E})=\mathbf{E}-\mathbf{C}_{w}(\mathbf{E})
\vspace{-0.2cm}
\end{equation}
where $\mathbf{C}_{w}(.)$ expresses the learned denoiser with weights $w$ satisfying RED conditions \cite{RED}, \cite{RED2} as well. It is noteworthy to mention that in (\ref{eq:25}) no computing of gradient is required and alternately, the denoiser network residual is easily exploited as the gradient which manifests the RED potential in terms of computational efficiency. Furthermore, RED provides us an explicit regularizer and consequently cost function in  (\ref{eq:20-1}) which encourages the convergence analysis and efficient tuning of hyper-parameters such as $\lambda$. To this end, the  explicit RED regularizer is expressed as:
\vspace{-0.3cm}
\begin{equation}
\label{eq:21}
R(\mathbf{E})_{RED}=\frac{1}{2}\left \langle \mathbf{E},\mathbf{E}-\mathbf{C}_{w}(\mathbf{E}) \right \rangle=\frac{1}{2}\mathbf{E}^{T}(\mathbf{E}-\mathbf{C}_{w}(\mathbf{E}))
\end{equation}
Benefiting the explicit objective function introduced by (\ref{eq:20}) and (\ref{eq:21}), the regularization parameter $\lambda$ can be tuned efficiently using the noise level information. In this regard, we utilize strong passivity condition of RED \cite{RED} which indicates that $\left \| \nabla _{\mathbf{E}} \mathbf{C}_{w}(\mathbf{E}) \right \|\leq 1$, thus, $\lambda$ at the equilibrium point can be estimated by:
\begin{equation}
\label{eq:26}
\lambda \geq \left \| \nabla g(\mathbf{E}) \right \|
\end{equation}
It is worth mentioning that the denoising network $\mathbf{C}_{w}$ is pre-trained supervisedly by feeding the clean elasticity images and the corrupted noisy ones $\tilde{\mathbf{E}}$ estimated by solving the minimization problem in (\ref{eq:20}) without any regularization as \textit{maximum likelihood} (ML) estimates of elasticity. \cmmnt{
using mean squared error (MSE) cost function as follows:
\vspace{-0.2cm}
\begin{equation}
\label{eq:22}
\emph{l}(w)=\frac{1}{2N}\sum^{N}\left \| C_{w}(\tilde{\mathbf{E}})-\mathbf{E} \right \|_{F}^{2}
\end{equation}
}
Once the denoiser network $\mathbf{C}_{w}$ is trained, the denoiser residual is fed into the GD estimation scheme in (\ref{eq:20-1}).\\
\begin{figure*}[t!]
\begin{minipage}[b]{0.19\linewidth}
  \centering
  \centerline{\includegraphics[width=3.0cm]{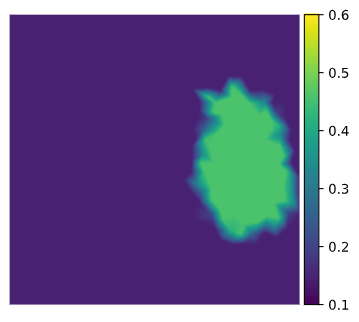}}
    \vspace{-0.6\baselineskip}
  \centerline{ \scriptsize{}}\medskip
\end{minipage}
\begin{minipage}[b]{.19\linewidth}
  \centering
  \centerline{\includegraphics[width=2.4cm]{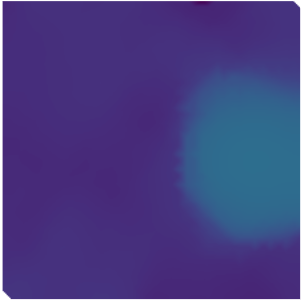}}
    \vspace{-0.6\baselineskip}
  \centerline{ \scriptsize{}}\medskip
\end{minipage}
\begin{minipage}[b]{0.19\linewidth}
  \centering
  \centerline{\includegraphics[width=2.4cm]{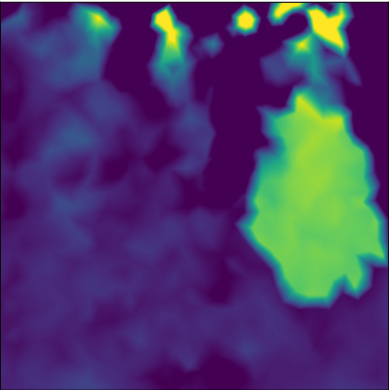}}
    \vspace{-0.6\baselineskip}
  \centerline{ \scriptsize{}}\medskip
\end{minipage}
\begin{minipage}[b]{0.19\linewidth}
  \centering
  \centerline{\includegraphics[width=2.4cm]{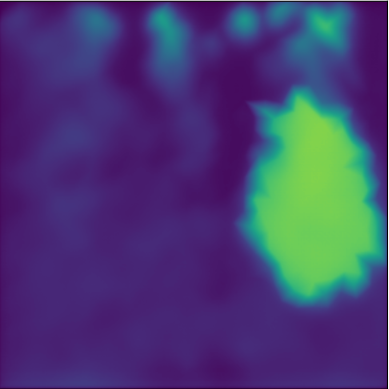}}
    \vspace{-0.6\baselineskip}
  \centerline{ \scriptsize{}}\medskip
\end{minipage}
\vspace{-0.35cm}
\caption{ Ground-truth and reconstructed elasticity images. (a) Ground-truth image. (b) Classical Gauss-Newton approach. (c) Statistical approach without regularization. (d) Proposed statistical approach with learned regularizarion by denoising (RED).  The unit of the color bar is 100 KPa. }
\vspace{-0.5cm}
\label{fig:2}
\end{figure*}
\begin{figure}[h]
\begin{minipage}[b]{0.8\linewidth}
  \centering
  \centerline{\includegraphics[width=5.8cm]{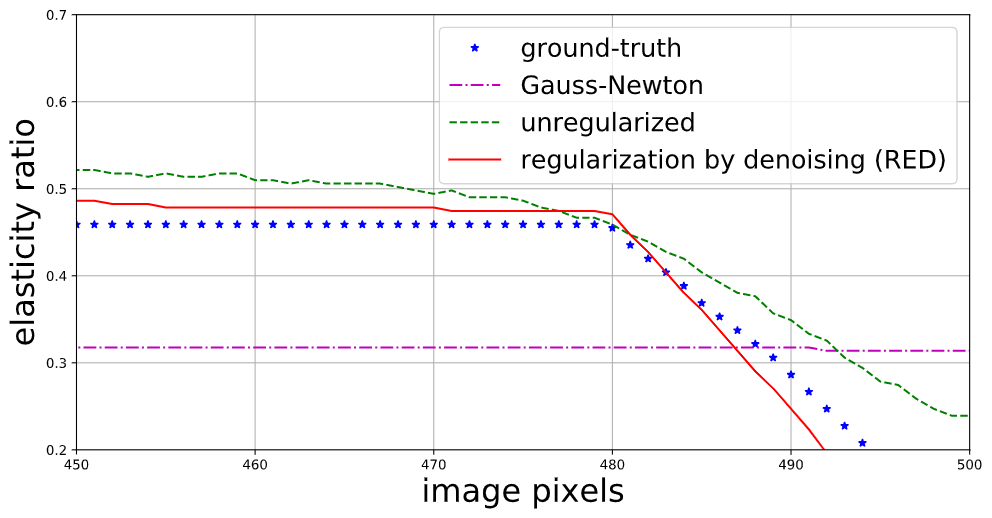}}
    \vspace{-0.4\baselineskip}
  \centerline{(a) \scriptsize{}}\medskip
\end{minipage}
\begin{minipage}[b]{.19\linewidth}
  \centering
  \centerline{\includegraphics[width=1.7cm]{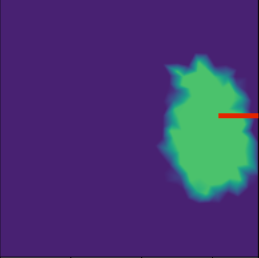}}
    \vspace{+0.4\baselineskip}
  \centerline{(b)\scriptsize{}}\medskip
\end{minipage}
  \vspace{-0.85cm}
\caption{  Cross-section comparison of reconstructed elasticity images indicated by the red line in (b).}
\label{fig:3}
\vspace{-0.55cm}
\end{figure}
Fig. \ref{fig:1} illustrates the overall elasticity estimation procedure. In this methodology, the initial elasticity fields $\mathbf{E}$ is a noisy ML elasticity estimate. Utilizing RED yields to explicitly separate regularizer learning from the data-fidelity term which expresses the statistical forward model of the imaging system. Following the fixed-point methodology, the forward model covariance matrix $\Gamma$ is estimated by feeding the current elasticity estimate $\mathbf{E}$ and then we compute the data-fidelity gradient
using the estimated $\Gamma$. Besides, the regularizer gradient is simply achieved by the denoiser network residual; and ultimately, these gradients are utilized in the GD update scheme. It must be noted that since the statistical forward model is exploited in the optimization problem explicitly, massive network weights are not wasted for the physical model learning which encourages the network training even with a limited training dataset.
\vspace{-0.2cm}
\section{Simulations and Results}
\label{sec5}
For performance evaluation, ultrasound elasticity images $\mathbf{E}$ are reconstructed using the measured noisy deformations $\mathbf{u^{m}}$ and Neumann BCs $\mathbf{f}$. In preliminary simulations, a dataset consisting of 540 mask images \cite{dataset} is used to generate ground-truth synthetic elasticity images representing some inclusions embedded in the background tissue. The normalized elasticity scale for each inclusion is selected in the range 0.3-0.8 KPa and the normalized elasticity scale for background tissue is selected in the range 0.1-0.15 KPa randomly. These settings lead to inclusion to background elasticity ratio in the range of 2-8 which follows experiments constraints. Using this dataset, the clean displacement images are generated by solving the deterministic global stiffness equation (\ref{eq:18}) and the noisy ones $\mathbf{u^{m}}$ are obtained by contaminating with multi-variate Gaussian noise $\mathbf{n}$ with $SNR=35dB$. UNet denoiser network $\mathbf{C}_{w}$ is trained using synthetic clean elasticity images and the corresponding noisy images. For generating noisy elasticity images, we map the noisy displacement fields $\mathbf{u^{m}}$ to the image domain by solving the unregularized inverse problem consisting of a data-fidelity term and a positivity constraint. The network residual is plugged into the reconstruction procedure depicted in Fig. \ref{fig:1} without any gradient computation.  Representative simulation results in Fig. \ref{fig:2} illustrate the effectiveness of the proposed approach for elasticity modulus reconstruction in presence of noise in comparison to conventional imaging using the Gaussian-Newton method as well as the statistical approach without learning-based regularization. For a detailed comparison of implemented approaches, cross-section representation of reconstructed elasticity fields $\mathbf{E}$ is depicted in Fig. \ref{fig:3} which manifests the RED method effectiveness.  
\vspace{-0.5cm}
\section{Conclusion}
\vspace{-0.2cm}
\label{sec6}
In this paper, we investigated a joint statistical and data-driven methodology for ultrasound elastography on the basis of FEMs by solving a regularized optimization problem. The proposed approach introduces an explicit integrated cost function including the statistical representation of the forward imaging model and a data-driven regularization term responsible for obtaining the underlying tissue elasticity distribution. The statistical representation of the forward model is implemented by a linear algebraic model with respect to the unknown elasticity image and a signal-dependent correlated model of noise. 
Following the RED paradigm, the data-driven regularizer is learned by supervisedly training a denoiser network and simply utilizing the network residual as the regularizer gradient which results in an explicit representation of the regularizer and subsequently the optimization problem. Leveraging theoretical convergence and uniqueness guarantees of the RED approach, the corresponding optimization task is solved by fixed-point iterative paradigm and GD which leads to stable elasticity estimation. Our simulation results verify the effectiveness of the proposed method. 
\vspace{-0.3cm}
\bibliographystyle{IEEEbib}
\bibliography{M335}

\end{document}